\documentclass[11pt]{article}
\pdfoutput=1

\usepackage{a4wide}
\usepackage{Packages}

\usepackage{Definitions}

\usepackage{comment}
\usepackage{setspace}

\newcommand{\OfficialTitle}{On the Perturbative Expansion around
  a Lifshitz Point}

\hypersetup{pdfauthor={Domenico Orlando, Susanne
    Reffert},pdftitle={\OfficialTitle}}

\author{
  \begin{minipage}{.97\linewidth}
    \vspace{1cm}
    \begin{center}
      \begin{small}
        \textbf{Domenico Orlando} and \textbf{Susanne Reffert}
      \end{small}
    \end{center}
    \vspace{1cm}% \hspace{2cm}
    \begin{center}
      \begin{minipage}{.7\linewidth}
        {\it \begin{footnotesize}
            \begin{center}
              Institute for the Mathematics and Physics of
              the Universe, \\The University of Tokyo, Kashiwa-no-Ha
              5-1-5, \\ Kashiwa-shi, 277-8568 Chiba, Japan.
            \end{center}
          \end{footnotesize}}
      \end{minipage}
    \end{center}
    \vspace{1cm}
  \end{minipage}
}

\date{}

\title{\vspace{1.5cm}
  \begin{huge}
    \textbf{\OfficialTitle}
  \end{huge}
}

\begin{document}

\numberwithin{equation}{section}
%\frontmatter

\begin{titlepage}
  \maketitle
  \thispagestyle{empty}

  \vspace{-13cm}
  \begin{flushright}
   IPMU-09-0105
  \end{flushright}

  \vspace{14cm}

  \begin{center}
    \textsc{Abstract}\\
  \end{center}

  The quantum Lifshitz model provides an effective description of a
  quantum critical point. It has been shown that even though
  non--Lorentz invariant, the action admits a natural
  supersymmetrization. In this note we introduce a perturbative
  framework and show how the supersymmetric structure can be used to
  greatly simplify the Feynman rules and thus the study of the model.

 \end{titlepage}

\onehalfspace

%\tableofcontents

%%%%%%%%%%%%%%%%%%%%%%%%%
\section{Introduction}
\label{sec:intro}

In 1941, Lifshitz~\cite{Lifshitz} introduced models with anisotropic
scaling between space and time in the context of tri--critical models.
Since then, such models have been studied in the context of solid
state physics. Materials with strongly correlated electrons, such as
copper oxides, show this type of critical behaviour, and also the
smectic phase of liquid crystals for example can be described this
way. Our treatment is based on quantum Lifshitz models as were studied
in~\cite{Ardonne:2003p1613}.  Quantum Lifshitz points are especially
interesting, since they are \emph{quantum critical
  points}~\cite{Sachdev}, \emph{i.e.} points at which a continuous
phase transition happens at $T=0$ which is driven by zero point
quantum fluctuations. A quantum Lifshitz point is characterized by the
vanishing of the term $(\nabla \phi)^2$ in the effective
Hamiltonian. While scale invariance is conserved, this gives rise to
an anisotropy between space and time.  This anisotropy is quantified
by the \emph{dynamical critical exponent} $z$,
\begin{equation} 
  t\to \lambda^zt,\ \ x\to \lambda x.
\end{equation}
For models in $2+1$ dimensions at a Lifshitz point, $z=2$, as opposed
to the Lorentz invariant $z=1$.

Models at a Lifshitz point have recently met with a large amount of
interest beyond their original field of application\footnote{We cannot
  give an extensive account of the existing literature here and thus
  content us to only mention a few main examples.}. A $3+1$
dimensional theory of gravity with $z=3$ put forward by
Ho\v rava~\cite{Horava:2009uw} has generated a big echo. But also in the
context of the AdS/CFT correspondence, interest in gravity duals of
non--Lorentz invariant models has arisen, see
\emph{e.g.}~\cite{Balasubramanian:2008dm,Son:2008ye,
  Kachru:2008yh,Volovich:2009yh}. In~\cite{Kachru:2008yh} in
particular, a gravity dual for a Lifshitz type model with $z=2$ was
proposed. As discussed recently in~\cite{Li:2009pf}, it seems
difficult to find string theory embeddings for gravity duals of
Lifshitz--type points.

While often, calculations are easier to do on the gravity side of the
correspondence, we are able to perform a number of calculations
directly on the field theory side, which are presented in this
article. Apart from being of interest directly for statistical
physics, our results can serve as a point of reference for comparison to results
derived on the gravity side.

\bigskip 

In~\cite{Dijkgraaf:2009gr} it was shown that systems of
Lifshitz type in $\left( d + 1 \right)$ dimensions admit a natural
supersymmetrization, a property which results from their relation to
$d$--dimensional models via a Langevin equation. The quantum Lifshitz
model in~\cite{Ardonne:2003p1613}, described by the
action 
\begin{align} 
  S [ \phi] = \int \diff \left[ \dot \phi^2 +
    (\partial_i \partial^i \phi )^2 \right], && \mathbf{x} = x_i, \, i
  = 1,2 \, ,
\end{align}
can be thought of as descending from a free boson in two dimensions with action
\begin{equation}
  W[\phi] = \int \di \mathbf{x} \, \left[ \partial_i \phi \partial^i \phi \right] \, .  
\end{equation}
This formulation allows the generalization of the quantum Lifshitz
model to massive and interacting cases. It becomes
possible to consider the class of models satisfying the detailed
balance condition whose (bosonic part of the) action takes the form
\begin{equation}
  S[\phi] = \int \diff \left[ \dot \phi^2 + \left( \frac{\delta W [\phi]}{\delta \phi }\right)^2 \right] \, ,
\end{equation}
where $W[\phi]$ is a local functional of the field $\phi
(t,\mathbf{x})$. The structure due to the Langevin
equation implies supersymmetry in the time direction, so that the
complete action includes also a fermionic field. It is given by
\begin{equation}
  \label{eq:SuperStochasticAction}
  S[\phi, \psi, \bar \psi] = \int \diff \left[ \dot \phi^2 + \left( \frac{\delta W [\phi]}{\delta \phi }\right)^2 - \bar \psi \left( \frac{\di}{\di t} + \frac{\delta^2 W[\phi]}{\delta \phi^2} \right) \psi \right] \, .
\end{equation}
This is the supersymmetric theory we focus on in this work.

A major advantage of models with this structure is that they can be
studied very efficiently by using a perturbative expansion of the
underlying Langevin equation, as proposed in~\cite{Cecotti:1983up}. In
this way, the cancellation of bosonic and fermionic terms in the
perturbative expansion becomes automatic. In consequence, there is a
\emph{great simplification of the Feynman diagrams} of the theory in
$\left( d+ 1 \right)$ dimensions which are reformulated in terms of
those of the $d$--dimensional system described by $W[\phi]$, plus a
set of additional rules. If we consider only $n$--point functions for
the bosonic field $\phi$, all the fermionic contributions are
automatically accounted for, so that it
is not even necessary to introduce a fermionic propagator.\\
For relativistic theories, this construction is possible only for
$d=0$ and $d=1$. Giving up Lorentz invariance, we concentrate on
$d=2$, which -- as we show in the following -- is the critical
case. The generalization to any $d$ is however clear.

\bigskip

In the following we derive
\begin{itemize}
\item the expression for the propagator of the free
  Lifshitz scalar (Sec. \ref{sec:propagator});
\item the Feynman rules for the simplest generalization to the
  interacting case (Sec. \ref{sec:interaction});
\item a scheme for UV regularization (Sec. \ref{sec:regularization}).
\end{itemize}
As examples, the three--point function
(Sec. \ref{sec:three-point-function}) and the one--loop propagator
(Sec. \ref{sec:one-loop-propagator}) are discussed.

%%%%%%%%%%%%%%%%%%%%%%%%%%%%%%%%
\section{The Langevin equation and the Nicolai map}
\label{sec:stoch-quant}

Having chosen to study the supersymmetric extension of the quantum
Lifshitz model, we can make use of the Nicolai map~\cite{Nicolai:1980jc}.  In a
supersymmetric field theory, a Nicolai map is a transformation of the
bosonic fields
\begin{equation}
\label{eq:Nicolai-map}
  \phi (t, \mathbf{x} ) \mapsto \eta(t, \mathbf{x} ) \, ,
\end{equation}
such that the bosonic part of the Lagrangian is quadratic in $\eta$
and the Jacobian for the transformation is given by the determinant of
the fermionic part:
\begin{gather}
  \label{eq:Nicolai-boson}
  S_B = \int \diff \left[ \frac{1}{2} \eta(t, \mathbf{x} )^2 \right] \, ,\\
  \label{eq:Nicolai-fermion}
  \det \left[ \frac{\delta \eta}{\delta \phi} \right] = \int \mathcal{D}\psi \mathcal{D} \bar \psi \, \exp[-S_F] \, .
\end{gather}
Following~\cite{Cecotti:1983up}, we would like to interpret the mapping
in Eq.~(\ref{eq:Nicolai-map}) as a Langevin equation for the field
$\phi(t, \mathbf{x})$ with noise $\eta(t,\mathbf{x})$. More precisely,
we want to show the equivalence of the action in Eq.~(\ref{eq:SuperStochasticAction})
 to the Langevin equation
\begin{equation}
  \label{eq:langevin_st}
  \frac{\partial\,\phi(t,\mathbf{x})}{\partial t} = -\frac{\delta  W}{\delta \phi}+\eta(t, \mathbf{x}).
\end{equation}
The correlations of $\eta$, which is a white Gaussian noise (as
in Eq.~(\ref{eq:Nicolai-boson})), are given by
\begin{align}
  \label{eq:corr_eta}
  \braket{\eta (t,\mathbf{x})} = 0 \, , && \braket{ \eta (t, \mathbf{x}))
    \eta(t^\prime, \mathbf{x}^\prime)} =  2\, \delta ( t - t^\prime)
  \delta ( \mathbf{x} - \mathbf{x}^\prime ) \, .
\end{align}
 A stochastic equation of
this type, where the dissipation term depends on the gradient of a function of the
field is said to satisfy the \emph{detailed balance condition}.
Equation~(\ref{eq:langevin_st}) has to be solved given an initial
condition, leading to an $\eta$--dependent solution
$\phi_\eta(t,\mathbf{x})$. As a consequence, $\phi_\eta(t,\mathbf{x})$ becomes a
stochastic variable. Its correlation functions are defined by
\begin{equation}
  \braket{\phi_\eta(t_1, \mathbf{x}_1)\ldots\phi_\eta(t_k, \mathbf{x}_k)}_\eta =
  \frac{\int\mathcal{D} \eta \, \exp \left[ -\tfrac{1}{4} \int \diff \eta^2(t,\mathbf{x}) \right] \phi_\eta(t_1,\mathbf{x}_1) \ldots
    \phi_\eta(t_k, \mathbf{x}_k)} {\int\mathcal{D}
    \eta \, \exp\left[-\tfrac{1}{4}\int \diff \eta^2(t,\mathbf{x})\right]} \, .
\end{equation}
A necessary condition for this approach to work is  that
thermal equilibrium is reached for $t\to \infty$, and that
\begin{equation}
  \label{eq:Equilibrium}
  \lim_{t\to\infty} \braket{\phi_\eta(t,\mathbf{x}_1) \ldots \phi_\eta(t,\mathbf{x}_k)}_\eta = \braket{\phi(\mathbf{x}_1) \ldots \phi(\mathbf{x}_k)},
\end{equation}
\emph{i.e.} that the equal time correlators for $\phi_\eta$ tend to
the corresponding quantum Green's functions for the $d$--dimensional
theory described by $W[\phi]$.  Since $\phi (t, \mathbf{x})$ is a
stochastic variable, the expectation value of any functional $F[\phi]$
is obtained by averaging over the noise:
\begin{equation}
  \braket{F[\phi]}_\eta = \frac{1}{\mathcal{Z}} \int \mathcal{D} \eta \,
  F[\phi] e^{-\frac{1}{2} \int \diff \eta
    (t,\mathbf{x})^2} \, ,
\end{equation}
where the partition function is defined by
\begin{equation}
  \mathcal{Z} = \int \mathcal{D} \eta \, e^{-\frac{1}{2} \int \diff \eta
    (t,\mathbf{x})^2} \, .
\end{equation}
It is convenient to change the integration variable from $\eta $ to
$\phi$.  The expression becomes
\begin{equation}
  \mathcal{Z} = \int \mathcal{D} \phi \, \left. \det \left[ \frac{\delta
        \eta}{\delta \phi} \right] \right|_{ \eta = \dot \phi + \frac{\delta W}{\delta \phi}} \exp \left[ -\frac{1}{2} \int \diff \left[  \left(\dot \phi +  \frac{\delta W}{\delta \phi} \right)^2  \right] \right] \, .
\end{equation}
The Jacobian can be expressed by introducing two fermionic fields
$\psi (t,\mathbf{x})$ and $\bar \psi(t,\mathbf{x})$ such that (as in
Eq.~(\ref{eq:Nicolai-fermion}))
\begin{equation}
  \left. \det \left[ \frac{\delta
        \eta}{\delta \phi} \right] \right|_{ \eta = \dot \phi + \frac{\delta W}{\delta \phi}}  = \int \mathcal{D} \psi \mathcal{D} \bar \psi
  \, \exp \left[ - \int \diff \left[\bar \psi(t,\mathbf{x}) \left( \partial_t -
      \frac{\delta^2 W}{\delta \phi^2}\right) \psi(t,\mathbf{x})  \right] \right] \, .
\end{equation}
In this way we can directly read off the $\left( d + 1
\right)$--dimensional action that, up to a boundary term, reproduces
Eq.~\eqref{eq:SuperStochasticAction}:
\begin{equation}
  S[\phi, \psi, \bar \psi] = \int \diff \left[ \dot \phi^2 + \left( \frac{\delta W [\phi]}{\delta \phi }\right)^2 - \bar \psi \left( \frac{\di}{\di t} + \frac{\delta^2 W[\phi]}{\delta \phi^2} \right) \psi \right] \, .
\end{equation}
Finally, one can show~\cite{Dijkgraaf:2009gr} that in a Hamiltonian formulation, the supersymmetric ground state is the bosonic state
\begin{equation}
	\ket{\Psi_0} = e^{-W[\phi]} \, ,
\end{equation}
as is already the case in the standard quantum Lifshitz model.

This construction bears an obvious resemblance with the stochastic
quantization of the $d$--dimensional theory described by $W[\phi]$. We
would however like to stress a fundamental difference. In the case of 
stochastic quantization, one is only interested in the $t \to
\infty$ limit and hence in the ground state. This means that the
action in Eq.~(\ref{eq:SuperStochasticAction}) is seen as a
topological theory. Here, on the other hand, we wish to study the
finite--time behaviour of the system, and the same action is taken to
describe a conventional supersymmetric model.

\bigskip

In the following, we will work in $d=2$, but in principle, all
calculations are equally valid for general $d$ and $z=2$.  The case
$d=2$ is arguably the most interesting, since it corresponds to the
Lifshitz point with its quantum critical behaviour.

%%%%%%%%%%%%%%%%%%%%%%%%%%%%%%%%%
\section{Perturbative Solution of the Langevin equation}
\label{sec:pert-solut-lang}

In the following, we will show how to perturbatively solve the Langevin equation (\ref{eq:langevin_st}), which gives rise to the dynamics of the
Lifshitz model. As we will see, the main advantage of this approach is
that the perturbative expansion is realized in terms of the Feynman
diagrams for the theory in $d$ dimensions which does not include
fermionic contributions.

To solve a transport equation like the Langevin equation in
Eq.\eqref{eq:langevin_st}, it is convenient to consider an integral
transform. The choice of transform depends on the choice of boundary
conditions. In space, the natural choice is given by requiring the
field to vanish at infinity,
\begin{equation}
  \phi (t, \mathbf{x} ) \xrightarrow[\abs{\mathbf{x}}\to \infty]{} 0 \, .  
\end{equation}
This means that the Fourier transform is well defined:
\begin{equation}
  \phi( t, \mathbf{k} ) = \int \di \mathbf{x} \, \left[ e^{- \imath \mathbf{k} \cdot \mathbf{x}} \phi (t, \mathbf{x} ) \right] \, .  
\end{equation}
For the time direction, we have two possible choices:
\begin{itemize}
\item[(a)] The field vanishes at \emph{negative infinity}. If we
  impose $\phi (t, \mathbf{x} ) \xrightarrow[t \to - \infty]{} 0 $, we
  can define a \emph{Fourier} transform in time and use
  \begin{equation}
    \phi ( \omega, \mathbf{k} ) = \int_{-\infty}^\infty \di t \, \left[e^{- \imath \omega t } \phi (t, \mathbf{k})  \right] \, .   
  \end{equation}
\item[(b)] Initial condition at $t = 0$. If we impose $\phi( 0,
  \mathbf{x} ) = \phi_0 (\mathbf{x})$, it is convenient to define a
  \emph{Laplace} transform in the time direction:
  \begin{equation}
    \phi ( s, \mathbf{k} ) = \int_0^\infty \di t \, \left[ e^{- s t } \phi (t,\mathbf{k})  \right] \, .    
  \end{equation}
\end{itemize}
Note that the first choice preserves time--translation invariance
which in case (b) is broken by an extra mode that describes the
evolution of the initial condition. On the other hand, if the kernel
of the Langevin equation $\frac{\delta W}{\delta \phi}$ is
\emph{positive definite} (as it is for the cases we are considering here),
this extra mode decays exponentially, and the large--time behaviours of
both choices coincide.  In other words, one can without loss of
generality choose to impose the initial condition (b) and then take
the large--time limit to recover the finite--time Fourier transform behaviour of
case (a)\footnote{An analogous problem was solved by Landau
  in~\cite{Landau:1946jc} in the study of oscillations in
  plasma.}. From now on, we will consider the Fourier--Laplace transform
(\emph{i.e.}  a Fourier transform in space and Laplace transform in the time direction):
\begin{equation}
  \phi (s, \mathbf{k} ) = \int_0^\infty \di t \int \di \mathbf{x} \left[ e^{- \imath \mathbf{k}\cdot \mathbf{x} - s t} \phi (t, \mathbf{x} ) \right] \, .  
\end{equation}

%%%%%%%%%%%%%%%
\subsection{Free propagator}
\label{sec:propagator}

As a first application, let us consider the action obtained by adding a relevant perturbation to the quantum Lifshitz model, described by
\begin{equation}
  S[ \phi, \psi, \bar \psi] =  \int \diff  \bigg[ \frac{1}{2} \dot \phi^2 +  ( \partial_i \partial^i \phi )^2 + m^2 \partial_i \phi\, \partial^i \phi + m^4 \phi^2 +\text{fermions} \bigg] \, .
\end{equation}

According to the argument in Sec.~\ref{sec:stoch-quant}, this is
equivalent to the Langevin equation corresponding to the massive boson
in $2$ dimensions described by the functional
\begin{equation}
    W [\phi] = \int \di \mathbf{x} \left[ \frac{1}{2}\partial_i \phi\, \partial^i \phi + \frac{1}{2} m^2 \phi^2   \right] \, .
\end{equation}
After the integral transform, the Langevin equation~(\ref{eq:langevin_st}) takes the form
\begin{equation}
  s\, \phi (s, \mathbf{k} ) - \phi_0 (\mathbf{k} ) = -  \Omega^2 \phi ( s, \mathbf{k} ) + \eta (s, \mathbf{k}) \, ,
\end{equation}
where we introduced $\Omega^2 =  \left( \mathbf{k}^2 + m^2 \right)
$. The Gaussian noise $\eta(s,\mathbf{k})$ has the
two--point function
\begin{equation}
  \braket{\eta (s,\mathbf{k}) \eta(s^\prime, \mathbf{k}^\prime) } = \frac{2 \left(2 \pi \right)^2 \delta (\mathbf{k} + \mathbf{k}^\prime)}{s + s^\prime} \, .   
\end{equation}
The retarded Green's function for this problem is the solution to the equation
\begin{equation}
  s\, G (s, \mathbf{k} ) = - \Omega^2 G( s, \mathbf{k} ) + 1.
\end{equation}
It follows that the solution to the Langevin equation is
\begin{equation}
  \phi (s, \mathbf{k} ) = G(s, \mathbf{k} ) \eta(s, \mathbf{k}) + G (s, \mathbf{k}) \phi_0 (\mathbf{k}) \, ,  
\end{equation}
and since the Laplace transform exchanges point--wise products and
convolution products, one finds that the field $\phi$ as a function of
time can be written as
\begin{equation}
  \phi(t, \mathbf{k}) = G(t, \mathbf{k}) \ast \eta( t, \mathbf{k}) + \phi_0 (\mathbf{k} ) G(t, \mathbf{k}) = \int_{0}^t \di \tau \, \left[  G( t - \tau, \mathbf{k}) \eta(\tau, \mathbf{k})  \right] + \phi_0 (\mathbf{k} ) G(t, \mathbf{k}) \, .
\end{equation}
More explicitly, using the fact that
\begin{equation}
  G(t, \mathbf{k} ) = e^{-t  \Omega^2 } \, ,
\end{equation}
we find the solution
\begin{equation}
  \phi(t, \mathbf{k}) = \int_{0}^t \di \tau \, \left[e^{- \left( t - \tau  \right)  \Omega^2 } \eta(\tau, \mathbf{k})  \right] + \phi_0 (\mathbf{k} ) e^{-t  \Omega^2 }  \, .
\end{equation}
Having expressed $\phi$ as a function of the noise, we are now in a
position to evaluate the two--point function:
\begin{multline}
  \label{eq:Propagator}
  D( s, \mathbf{k} ; s^\prime, \mathbf{k}^\prime ) = \braket{ \phi( s, \mathbf{k} ) \phi (s^\prime,\mathbf{k}^\prime) } = G( s, \mathbf{k} ) G( s^\prime, \mathbf{k}^\prime) \braket{ \eta( s, \mathbf{k} ) \eta (s^\prime,\mathbf{k}^\prime) } \\ = \frac{2 \left(2 \pi \right)^2 \delta( \mathbf{k} + \mathbf{k}^\prime)}{\left( s +  \Omega^2 \right) \left( s^\prime +  \Omega^2 \right) \left( s + s^\prime \right)} \, .
\end{multline}
Taking the (bidimensional) inverse Laplace transform, this becomes
\begin{multline}
  D( t, \mathbf{k}; t^\prime, \mathbf{k}^\prime ) = \int_{\sigma -
    \imath \infty}^{\sigma + \imath \infty} \frac{\di s}{2 \pi \imath}
  \int_{\sigma - \imath \infty}^{\sigma + \imath \infty} \frac{\di
    s^\prime}{2 \pi \imath} \, \left[ e^{s t + s^\prime t^\prime}D( s,
  \mathbf{k} ; s^\prime, \mathbf{k}^\prime ) \right] \\ = \left( 2 \pi
  \right)^2 \frac{e^{- \Omega^2 \abs{t - t^\prime}} - e^{- \Omega^2
      \left( t + t^\prime\right)}}{ \Omega^2} \delta ( \mathbf{k} +
  \mathbf{k}^\prime ) \, .
\end{multline}
Two limits are interesting:
\begin{itemize}
\item For $t, t^\prime \to \infty$, the second exponential is
  suppressed and the two--point function becomes
  \begin{equation}
    D( t, \mathbf{k}; t^\prime, \mathbf{k}^\prime ) \xrightarrow[t, t^\prime \to \infty]{} \left(2 \pi \right)^2 \frac{e^{-  \Omega^2 \abs{t - t^\prime}}}{ \Omega^2} \delta ( \mathbf{k} + \mathbf{k}^\prime ) \, .
  \end{equation}
  Its Fourier transform is given by
  \begin{equation}
    D ( \omega, \mathbf{k}; \omega^\prime, \mathbf{k}^\prime ) = \frac{\delta (\omega + \omega^\prime) \delta( \mathbf{k} + \mathbf{k}^\prime)}{\omega^2 + \Omega^2} \, ,
  \end{equation}
  as found in~\cite{Ardonne:2003p1613}.
\item For $t = t^\prime \to \infty$, the two--point function reproduces
  the usual bosonic propagator in $d$ dimensions (as expected from the
  general structure of the Langevin equation and shown in
  Eq.~(\ref{eq:Equilibrium})):
  \begin{equation}
    D( t, \mathbf{k}; t, \mathbf{k}^\prime ) \xrightarrow[t \to \infty]{} \left(2 \pi \right)^2 \frac{\delta (\mathbf{k} + \mathbf{k}^\prime)}{ \Omega^2} \, .
  \end{equation}
 Note that for $m = 0$, this means that for large times, the
 propagator will behave polynomially and the theory is critical, just
  like in the case of a two--dimensional bosonic field.
\end{itemize}

%%%%%%%%%%%%%%%%%%%%%%%%%%%%%%%%%%%%%%%%%%%%%
\subsection{Interacting theory}
\label{sec:interaction}

As an example of an interacting theory, let us consider the action
descending from the theory of a massive boson with a $\phi^3 $
interaction. This is described by a Langevin equation with functional
\begin{equation}
  W [\phi] = \int \di \mathbf{x} \left[ \frac{1}{2}\partial_i \phi\, \partial^i \phi + \frac{1}{2} m^2 \phi^2 + \frac{1}{3} g^2 \phi^3  \right] .
\end{equation}
The action in $\left(2 + 1 \right)$ dimensions in
Eq.~(\ref{eq:SuperStochasticAction}) is immediately found
to be
\begin{multline}
  \label{eq:InteractingMassiveAction}
  S[ \phi, \psi, \bar \psi] =  \int \diff  \bigg[ \frac{1}{2} \dot \phi^2 +  ( \partial_i \partial^i \phi )^2 + m^2 \partial_i \phi\, \partial^i \phi + g^2 \phi \,\partial_i \phi\, \partial^i \phi + m^4 \phi^2 + g^2 m^2 \phi^3 + g^4 \phi^4 +  \\ +  \text{fermions} \bigg] \, ,
\end{multline}
and in particular, for the critical $m=0$ case, 
\begin{equation}
  S[ \phi] = \int \diff \left[ \frac{1}{2} \dot \phi^2 + ( \partial_i \partial^i \phi )^2 + g^2 \phi\, \partial_i \phi \,\partial^i \phi +  g^4 \phi^4 + \text{fermions} \right] \, .
\end{equation}
Note that the coefficients of the three-- and four--point interactions
are not independent, but fixed by the detailed balance condition.  It
is in fact the detailed balance, which here plays the the role of a
symmetry the theory has to fulfill, which keeps terms other than those
given in Eq. (\ref{eq:InteractingMassiveAction}) from appearing.  This
relation between the different coupling constants is a property which
is accessible to experimental checks in materials which are described
by a Lifshitz point effective action and thus is a testable prediction
of this framework.

The most effective way of making use of this symmetry consists once
more in starting from the corresponding Langevin equation, which takes
the simple form
\begin{equation}
  \partial_t \phi (t, \mathbf{x} ) = -  \left( - \nabla^2 \phi (t,\mathbf{x})  + m^2 \phi(t,\mathbf{x}) + g^2 \phi^2(t,\mathbf{x}) \right) + \eta (t,\mathbf{x}) \, ,
\end{equation}
where $\eta (t, \mathbf{x})$ is the same Gaussian noise as in the
previous example.  Taking the Fourier--Laplace transform as above, one
has to deal with the quadratic term that can be recast in the
following form:
\begin{multline}
  -g ^2 \int_0^\infty \di t \int \di\mathbf{x} \, e^{-\imath \mathbf{k}
    \cdot \mathbf{x} - s t} \phi ( t, \mathbf{x} )^2 =\\
  = - g^2\int \diff \frac{\di \mathbf{k}_1}{\left(2 \pi \right)^2}
  \frac{\di \mathbf{k}_2}{\left(2 \pi \right)^2} \frac{\di s_1}{2 \pi
    \imath}\frac{\di s_2}{2 \pi \imath} \, \left[ e^{-\imath \left(
      \mathbf{k}- \mathbf{k}_1 - \mathbf{k}_2 \right) \cdot \mathbf{x}
    - \left( s - s_1 - s_2 \right) t} \phi ( s_1, \mathbf{k}_1 ) \phi
  ( s_2, \mathbf{k}_2 )  \right] =  \\
  = - g^2\int d[\mathbf{k_1},\mathbf{k}_2,s_1,s_2] \, \left[ \frac{\delta(\mathbf{k} -
    \mathbf{k}_1 - \mathbf{k}_2)}{s - s_1 - s_2} \phi ( s_1,
  \mathbf{k}_1 ) \phi ( s_2, \mathbf{k}_2 )  \right]\, .
\end{multline}
The Langevin equation becomes an integral equation:
\begin{multline}
  \phi (s, \mathbf{k} ) = G(s, \mathbf{k} ) \eta(s, \mathbf{k}) - g^2 G(s, \mathbf{k} )
  \int \di [\mathbf{k}_1, \mathbf{k}_2, s_1, s_2] \left[
    \frac{\delta(\mathbf{k} - \mathbf{k}_1 - \mathbf{k}_2)}{s - s_1 -
      s_2} \phi ( s_1, \mathbf{k}_1 ) \phi ( s_2, \mathbf{k}_2 )
  \right] \\
  + G (s, \mathbf{k})\, \phi_0 (\mathbf{k})\, .
\end{multline}
This type of equation can be solved perturbatively in $g^2$, using the
usual Feynman diagram techniques by denoting $G (s, \mathbf{k})$ with
an arrow and the noise $\eta( s, \mathbf{ k})$ with a cross. In
particular, the field $\phi$ can be expanded as
\begin{equation}
\label{eq:FieldExpansion}
  \phi =\includegraphics[scale=.7]{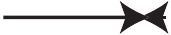}
+
\begin{minipage}{.20\linewidth}
  \includegraphics[scale=.7]{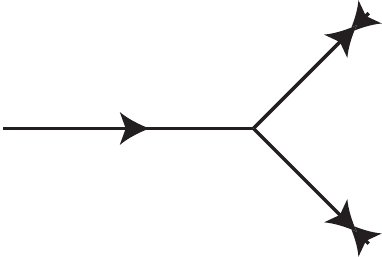}
\end{minipage}
+
\begin{minipage}{.25\linewidth}
  \includegraphics[scale=.7]{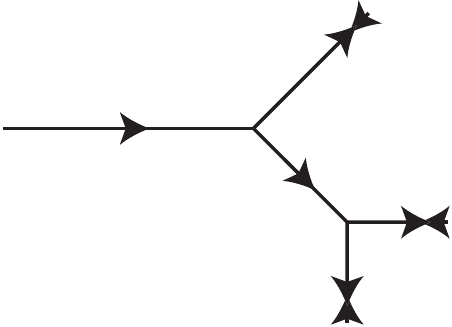}
\end{minipage}
+ \dots
\end{equation}
The Feynman rules for this model are summarized in
Table~\ref{tab:FeynmanRules}.
\begin{table}
  \centering
  \begin{tabular}{c@{\extracolsep{2em}}cc}
    \toprule
    element & Fourier-Laplace & Fourier \\ \midrule
    \begin{minipage}{.14\linewidth}
      \begin{picture}(20,5)(,)
        \begin{small}
          \put(22,10){$s$} \put(22,-7){$\mathbf{k}$}
        \end{small}
        \includegraphics[scale=.7]{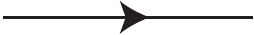}
      \end{picture}
    \end{minipage}
    & $G(s, \mathbf{k}) = \displaystyle{\frac{1}{s + \Omega^2}}$ & $G(\omega, \mathbf{k}) = \displaystyle{\frac{1}{\imath \omega + \Omega^2}}$ \\[.5cm] 
    \begin{minipage}{.14\linewidth}
      \begin{picture}(20,5)(,)
        \begin{small}
          \put(2,10){$s$} \put(2,-7){$\mathbf{k}$}
          \put(54,10){$s^\prime$} \put(54,-7){$\mathbf{k}^\prime$}
        \end{small}
        \includegraphics[scale=.7]{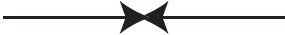}
      \end{picture}
    \end{minipage}& $ \displaystyle{\frac{2 \left(2 \pi \right)^2 \delta( \mathbf{k} + \mathbf{k}^\prime)}{\left( s +  \Omega^2 \right) \left( s^\prime +  \Omega^2 \right) \left( s + s^\prime \right)}} $ & $\displaystyle{\frac{\left(2 \pi \right)^{2+1} \delta (\mathbf{k} + \mathbf{k}^\prime) \delta( \omega + \omega^\prime)}{\omega^2 + \Omega^4}}$ \\[.7cm]
    \begin{minipage}{.15\linewidth}
      \begin{picture}(20,30)(,) 
        \begin{small}
          \put(0,20){$s_1$} \put(0,2){$\mathbf{k}_1$} 
          \put(30,30){$s_2
            \hspace{.6em}\mathbf{k}_2$} \put(30,-4){$s_3
            \hspace{.6em}\mathbf{k}_3$}
        \end{small}
        \includegraphics[scale=.7]{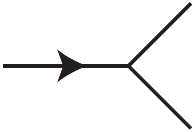}
      \end{picture}
    \end{minipage}
& $g^2 \displaystyle{\frac{\delta( \mathbf{k}_1 - \mathbf{k}_2 - \mathbf{k}_3)}{s_1 - s_2 - s_3}} $ & $g^2 \delta( \mathbf{k}_1 - \mathbf{k}_2 - \mathbf{k}_3) \delta(\omega_1 - \omega_2 - \omega_3)$ \\[.7cm] \bottomrule
  \end{tabular}
  \caption{Feynman rules for the cubic theory}
  \label{tab:FeynmanRules}
\end{table}
Note that even if the action in
Eq.~\eqref{eq:InteractingMassiveAction} has two cubic and a quartic
interaction, the Feynman diagrams obtained from the Langevin equation
only have one cubic vertex.  This might at first seem surprising, but is again a simplifying consequence of the detailed balance condition.

%%%%%%%%%%%%%%%%%%%%%%%%%%%%

\subsection{Examples}

\subsubsection{Three--point function}
\label{sec:three-point-function}

As a first example, let us consider the three--point function
$\braket{\phi(s_1, \mathbf{k}_1)\phi(s_2, \mathbf{k}_2)\phi(s_3,
  \mathbf{k}_3)}$ at tree level. Expanding the field as in
Eq.~\eqref{eq:FieldExpansion}, we find that at tree level the function
is given by the sum of three contributions:
\begin{equation}
  \braket{\phi(s_1, \mathbf{k}_1)\phi(s_2, \mathbf{k}_2)\phi(s_3,
    \mathbf{k}_3)} \hspace{.5em} = \hspace{1em}
  \begin{small}
    \begin{minipage}{.08\linewidth}
      \begin{picture}(20,37)(,) 
        \put(0,26){$s_1$}
        \put(0,8){$\mathbf{k}_1$}
        \put(18,40){$s_2 \hspace{.6em}\mathbf{k}_2$}
        \put(18,-4){$s_3 \hspace{.6em}\mathbf{k}_3$}
        \includegraphics[scale=.5]{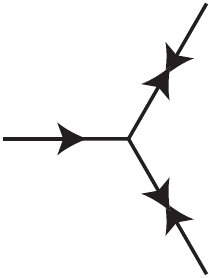}
      \end{picture}
    \end{minipage}
    \hspace{1em} + \hspace{1em} \begin{minipage}{.08\linewidth}
      \begin{picture}(20,38)(,) 
        \put(0,26){$s_1$}
        \put(0,8){$\mathbf{k}_1$}
        \put(22,40){$s_2 \hspace{.6em}\mathbf{k}_2$}
        \put(22,-4){$s_3 \hspace{.6em}\mathbf{k}_3$}
      \includegraphics[scale=.5]{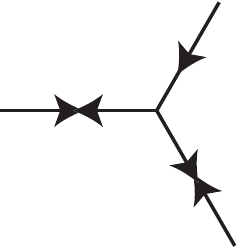}
    \end{picture}
  \end{minipage} \hspace{1em} + \hspace{1em}
  \begin{minipage}{.08\linewidth}
      \begin{picture}(20,30)(,) 
        \put(0,24){$s_1$}
        \put(0,6){$\mathbf{k}_1$}
        \put(22,40){$s_2 \hspace{.6em}\mathbf{k}_2$}
        \put(22,-4){$s_3 \hspace{.6em}\mathbf{k}_3$}
      \includegraphics[scale=.5]{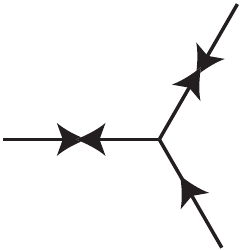}
      \end{picture}
    \end{minipage}
  \end{small}
\end{equation}
Using the Feynman rules in Table~\ref{tab:FeynmanRules}, it is
immediate to find that each diagram gives a contribution of
\begin{equation}
  -\int \frac{\di s_a \di s_b}{4\pi^2} \left[ \textstyle{\frac{4 g^2}{\left( s_i + \Omega_i^2 \right) \left( s_j + \Omega_j^2 \right) \left( s_a + \Omega_j^2 \right) \left( s_j + s_a \right) \left( s_k + \Omega_k^2 \right) \left( s_b + \Omega_k^2 \right) \left( s_k + s_b \right) \left( s_i - s_a - s_b\right)}}  \right] \, ,
\end{equation}
where $\set{i,j,k}$ take the values $\set{1,2,3}$ and their cyclic
permutations. After an inverse Laplace transform in the time direction
and taking the large time limit (which together is equivalent to
taking the Fourier transform), the three--point function becomes:
\begin{multline}
  \braket{\phi(t_1, \mathbf{k}_1)\phi(t_2, \mathbf{k}_2)\phi(t_3, \mathbf{k}_3)} \\
  = g^2\tfrac{e^{-\Omega_2^2 t_{21}- \Omega_3^2 t_{31} \Omega_1^2 \left(
        \Omega_1^2+ \Omega_2^2 - \Omega_3^2 \right)} + e^{-\Omega_1^2
      t_{31} - \Omega_2^2 t_{32}} \Omega_3^2 \left( - \Omega_1^2 +
      \Omega_2^2 + \Omega_3^2 \right) - e^{-\Omega_1^2 t_{21} -
      \Omega_3^2 t_{32}} \Omega_2^2 \left( \Omega_1^2 + \Omega_2^2 +
      \Omega_3^2 \right)}{\Omega_1^2 \Omega_2^2 \Omega_3^2 \left(
      \left( \Omega_1^2 - \Omega_3^2 \right)^2 - \Omega_2^4 \right)} ,
\end{multline}
where $t_{ij} = t_i - t_j$ and $t_1 \le t_2 \le t_3$. A consistency
check for this expression is obtained by considering the equal time
case $t_1 = t_2 = t_3 \to \infty$, which reproduces, as expected, the
usual bosonic result (as in Eq.~(\ref{eq:Equilibrium})):
\begin{equation}
  \braket{\phi(t, \mathbf{k}_1)\phi(t, \mathbf{k}_2)\phi(t, \mathbf{k}_3)} \xrightarrow[t \to \infty]{} \frac{g^2}{\Omega_1^2 \Omega_2^2 \Omega_3^2}\,.
\end{equation}

%%%%%%%%%%%%%%%%%%%%%%
\subsubsection{One--loop propagator}
\label{sec:one-loop-propagator}

As a next example, let us consider the one--loop correction to the
two--point function $\braket{\phi(s_1, \mathbf{k}_1) \phi( s_2,
  \mathbf{k}_2 )}$. Using the expansion in
Eq.~\eqref{eq:FieldExpansion}, we find that
\begin{equation}
  \braket{\phi(s_1, \mathbf{k}_1) \phi(s_2, \mathbf{k}) } = 
  \begin{small}
    \begin{minipage}{.14\linewidth}
      \vspace{3em}
      % TeXnote: the vspace are put so that all the labels on the
      % graphs fit in the equation 
      \begin{picture}(20,5)(,)
        \put(2,10){$s_1$}
        \put(2,-7){$\mathbf{k}$}
        \put(47,10){$s_2$}
        \put(47,-7){$\mathbf{k}$}
        \put(33,-22){(a)}
        \includegraphics[scale=.7]{StochGraph-Propagator-crop}
      \end{picture}
      \vspace{3em}
    \end{minipage} +
    \begin{minipage}{.17\linewidth}
      \begin{picture}(20,30)(,) 
        \put(2,22){$s_1$}
        \put(2,5){$\mathbf{k}$} 
        \put(66,22){$s_2$}
        \put(66,5){$\mathbf{k}$}
        \put(18,30){$s_a$}
        \put(18,0){$s_b$}
        \put(48,30){$s_d$}
        \put(48,0){$s_c$}
        \put(33,35){$\mathbf{k}_1$}
        \put(33,-8){$\mathbf{k}_2$}
        \put(33,-22){(b)}
        \includegraphics[scale=.5]{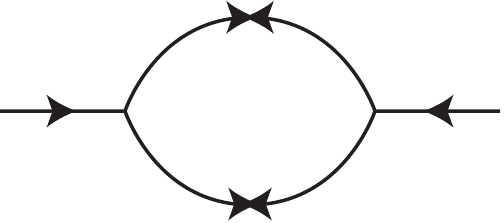}
      \end{picture}
    \end{minipage} +
    \begin{minipage}{.17\linewidth}
      \begin{picture}(20,30)(,) 
        \put(0,22){$s_1$}
        \put(0,5){$\mathbf{k}$} 
        \put(66,22){$s_2$}
        \put(66,5){$\mathbf{k}$}
        \put(12,22){$s_a$}
        \put(18,30){$s_b$}
        \put(18,0){$s_c$}
        \put(48,30){$s_d$}
        \put(33,35){$\mathbf{k}_1$}
        \put(33,-8){$\mathbf{k}_2$}
        \put(33,-22){(c)}
       \includegraphics[scale=.5]{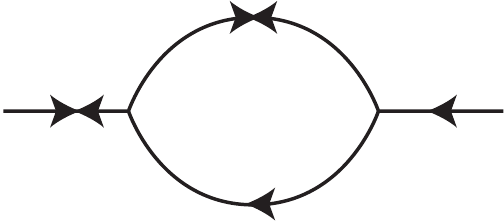}
     \end{picture}
    \end{minipage} +
    \begin{minipage}{.17\linewidth}
      \begin{picture}(20,30)(,) 
        \put(0,22){$s_1$}
        \put(0,5){$\mathbf{k}$} 
        \put(66,22){$s_2$}
        \put(66,5){$\mathbf{k}$}
        \put(52,22){$s_d$}
        \put(18,30){$s_a$}
        \put(18,0){$s_b$}
        \put(46,30){$s_c$}
        \put(33,35){$\mathbf{k}_1$}
        \put(33,-8){$\mathbf{k}_2$}
        \put(33,-22){(d)}
      \includegraphics[scale=.5]{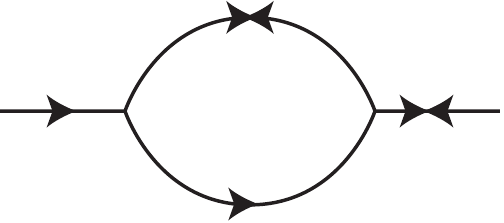}
    \end{picture}
  \end{minipage}
  \end{small}
\end{equation}
The first term is just the usual propagator $D( s_1, \mathbf{k}_1;
s_2, \mathbf{k}_2)$. The contributions from the diagrams (b), (c), (d)
are as follows:
\begin{align}
  \text{(b)} &= g^4\frac{G(s_1, \mathbf{k}) D(s_a,\mathbf{k}_1; s_d, -\mathbf{k}_1 ) D( s_b, \mathbf{k}_2; s_c, -\mathbf{k}_2) G( s_2, \mathbf{k})}{\left( s_1-s_a-s_b \right) \left( s_2 - s_d - s_c \right)} \, , \\
  \text{(c)} &= g^4\frac{D( s_1, \mathbf{k}; s_a, - \mathbf{k} ) D( s_b, \mathbf{k}_1; s_d, -\mathbf{k}_1) G (s_c, \mathbf{k_2} ) G( s_2, \mathbf{k} )}{\left( s_c - s_a - s_b \right) \left( s_2 - s_c - s_d \right)} \, ,\\
  \text{(d)} &= g^4\frac{G(s_1, \mathbf{k}) D(s_a,\mathbf{k}_1; s_c,
    -\mathbf{k}_1) G(s_b,\mathbf{k}_2) D(s_d, \mathbf{k}; s_2;
    -\mathbf{k}) }{\left( s_1 - s_a - s_b \right) \left( s_b - s_c -
      s_d \right)} \, .
\end{align}
The two--point function is obtained by summing up all the contributions
and integrating over the internal momenta.  Once more, the result is
more transparent if we take the inverse Laplace transform and consider
the large time limit $t_1, t_2 \to \infty$:
\begin{multline}
  \braket{\phi(t_1, \mathbf{k}) \phi(t_2, \mathbf{k}) }
  \xrightarrow[t_1, t_2 \to \infty]{} \frac{e^{-\Omega^2
      t_{21}}}{\Omega^2} \\
  + g^4 \int \frac{\di \mathbf{k}_1 \di \mathbf{k}_2}{\left(2 \pi
    \right)^{4}} \left[ \frac{\Omega^2 e^{- \left( \Omega_1^2 +
          \Omega_2^2 \right) t_{21}} - \left( \Omega_1^2 + \Omega_2^2
      \right) e^{-\Omega^2 t_{21}}}{\Omega^4 \Omega_1^2 \Omega_2^2
      \left( \Omega^2 - \Omega_1^2 - \Omega_2^2 \right)} \delta (
    \mathbf{k} - \mathbf{k}_1 - \mathbf{k}_2 ) \right] \, ,
\end{multline}
where $t_{21} = \abs{t_2 - t_1}$. Taking the $t_{21} \to 0$ limit, one
reproduces the usual bosonic two--point function at one loop (as in
Eq.~(\ref{eq:Equilibrium})):
\begin{equation}
  \braket{\phi(t, \mathbf{k}) \phi(t, \mathbf{k}) } \xrightarrow[t\to \infty]{} \frac{1}{\Omega^2} +  g^4 \int \frac{\di \mathbf{k}_1 \di \mathbf{k}_2}{\left(2 \pi \right)^{4}} \frac{ \delta (\mathbf{k} - \mathbf{k}_1 - \mathbf{k}_2) }{\Omega^4 \Omega_1^2 \Omega_2^2}\, .
\end{equation}

%%%%%%%%%%%%%%%%%%%%%%%%%
\subsection{UV Regularization}
\label{sec:regularization}

The two--point function we derived above suffers from an ultraviolet
divergence. In order to regularize it, one can either change the
Langevin equation (Eq.~(\ref{eq:langevin_st})), or the noise
correlation function in Eq.~(\ref{eq:corr_eta}). Here, we follow the
latter approach and show how smearing out the delta function
\emph{in time} by introducing a cut--off $\Lambda$ results in an
ultraviolet regularization \emph{in space} which, in the large time
limit, reproduces the usual Pauli--Villars result.

Consider the noise function $\eta_\Lambda(t, \mathbf{x})$ with the
following two--point function:
\begin{equation}
  \braket{\eta_\Lambda (t, \mathbf{x}) \eta_\Lambda (t^\prime, \mathbf{x}^\prime)} = \delta (\mathbf{x} - \mathbf{x}^\prime ) \Lambda^2 e^{-\Lambda^2 \abs{t - t^\prime}} \, .
\end{equation}
For $\Lambda \to \infty$, it converges to the two--point function of the usual noise:
\begin{gather}
  \braket{\eta_\Lambda (t, \mathbf{x}) \eta_\Lambda (t^\prime, \mathbf{x}^\prime)} \xrightarrow[\Lambda \to \infty]{} 2  \delta (\mathbf{x} - \mathbf{x}^\prime ) \delta (t - t^\prime ) \,,\\
  \eta_{\Lambda} (t, \mathbf{x} ) \xrightarrow[\Lambda \to \infty]{} \eta (t, \mathbf{x}) \, .
\end{gather}
Applying the Fourier--Laplace transform, we get
\begin{equation}
  \braket{\eta_\Lambda (t, \mathbf{x}) \eta_\Lambda (t^\prime, \mathbf{x}^\prime)} = \frac{ \left( 2\pi \right)^2 \delta (\mathbf{k} + \mathbf{k}^\prime )}{ s + s^\prime} \frac{ \Lambda^2  \left(  s + s^\prime + 2 \Lambda^2 \right) }{\left( s + \Lambda^2 \right) \left( s^\prime + \Lambda^2 \right)} \, .
\end{equation}
In this scheme, the Langevin equation remains unchanged, which implies
that the retarded Green's function remains the same,
\begin{equation}
  G ( s, \mathbf{k} )  = \frac{1}{s +  \Omega^2} \, ,  
\end{equation}
while the field receives a $\Lambda^2 $ correction. In particular, for
the free case:
\begin{equation}
  \phi_\Lambda (s, \mathbf{k} ) = G( s, \mathbf{k} ) \eta_{\Lambda} (s, \mathbf{k}) + G( s, \mathbf{k}) \phi_0 (\mathbf{k}) \, .  
\end{equation}
We are now in a position to calculate the $\Lambda^2 $ correction to
the propagator in Eq.~(\ref{eq:Propagator}):
\begin{equation}
  D_{\Lambda} (s, \mathbf{k}; s^\prime, \mathbf{k}^\prime ) = \braket{\phi_{\Lambda}(s, \mathbf{k}) \phi_{\Lambda}( s^\prime, \mathbf{k}^\prime) } = \frac{ \left(2 \pi \right)^2 \delta( \mathbf{k} + \mathbf{k}^\prime)}{\left( s +  \Omega^2 \right) \left( s^\prime +  \Omega^2 \right) \left( s + s^\prime \right)}  \frac{ \Lambda^2  \left(  s + s^\prime + 2 \Lambda^2 \right) }{\left( s + \Lambda^2 \right) \left( s^\prime + \Lambda^2 \right)} \, .  
\end{equation}
To see how this corresponds to an ultraviolet regularization, let us
perform the inverse Laplace transform:
\begin{multline}
  D_{\Lambda} (t, \mathbf{k}; t^\prime, \mathbf{k}^\prime ) = \frac{ \left( 2\pi \right)^2 \delta (\mathbf{k} + \mathbf{k}^\prime ) \Lambda^2}{\Omega^2 \left( \Lambda^4 - \Omega^4 \right)} \left( \Lambda^2 \ e^{- \Omega^2 \abs{t - t^\prime}}  - \Omega^2 \ e^{- \Lambda^2 \abs{t - t^\prime}} \right. \\ \left.- \left( \Lambda^2 + \Omega^2 \right) e^{-\Omega^2 \left( t + t^\prime \right)} + \Omega^2 e^{-\Omega^2 t - \Lambda^2 t^\prime} + \Omega^2 e^{-\Omega^2 t^\prime - \Lambda^2 t} \right) \, .
\end{multline}
For large times, only the first two terms in the sum contribute, and
for $t = t^\prime \to \infty$, we find the usual Pauli--Villars
propagator for the $2$--dimensional boson:
\begin{equation}
  D( t, \mathbf{k}; t, \mathbf{k}^\prime ) \xrightarrow[t \to \infty]{} \left(2 \pi \right)^2 \delta (\mathbf{k} + \mathbf{k}^\prime)\frac{\Lambda^2}{\Omega^2 \left( \Lambda^2 + \Omega^2 \right)} \, .
\end{equation}

%%%%%%%%%%%%%%%%%%%%%%%%%%%%%
\subsection*{Acknowledgements}

We are indebted to Luciano Girardello for
enlightening discussions, and Tadashi Takayanagi for illuminating 
discussions and comments on the manuscript. We would also like to thank Masahito
Yamazaki for numerous discussions at an early stage of the project. Finally, we would
like to thank the workshop ``New Perspectives in String Theory'' at
the Galileo Galilei Institute for Theoretical Physics in Arcetri,
Firenze, for hospitality, where this research was initiated.

The research of the authors was supported by the World Premier
International Research Center Initiative (WPI Initiative), MEXT,
Japan.

\bibliography{References}

\end{document}